\documentclass[twocolumn,showpacs,preprintnumbers,amsmath,amssymb]{revtex4}
\usepackage{amssymb}
\usepackage[dvips]{graphicx}
\begin{document}
\title { Condensed phase of  converting boson-fermion mixtures}

\author{A. S. Alexandrov}
\affiliation {Department of Physics, Loughborough University,
Loughborough LE11 3TU, United Kingdom}

\begin{abstract}

\noindent Theory of a condensed state of hybridised bosons and
fermions is developed.  Normal and anomalous
  Green's functions are obtained diagrammatically and analytically using the Hamiltonian
  of the boson-fermion model (BFM). A pairing of
 bosons analogous to the Cooper pairing of
 fermions is found.
There are three coupled condensates in the model, described by the
 off-diagonal  single-particle boson,  pair-fermion and pair-boson
 fields.
  The Gor'kov expansion in the strength of the order
parameter near the transition yields no linear homogeneous term in
the Ginzburg-Landau equation, and the \emph{infinite}
Levanyuk-Ginzburg parameter, $Gi=\infty$, which indicates that
previous mean-field discussions  of BFM are flawed.

\end{abstract}

\maketitle

A two component model of  negative $U$ centers coupled with the
Fermi sea of itinerant fermions was originally employed to study
superconductivity in disordered metal-semiconductor alloys
\cite{sim,tin}. Later on it was applied more generally to describe
pairing electron processes with localisation-delocalisation
\cite{ion},  and to the polaron-bipolaron crossover problem in the
intermediate electron-phonon coupling regime \cite{rob}. The model
attracted more attention  in connection with exotic \cite{rob2},
and high-temperature superconductors
\cite{eli,lee,ran,ran0,kos,lar,ale,ale2,ren,cris,gor,dam,mic}.
When the attractive potential $U$ is large, the model is reduced
to localised bosons  spontaneously decaying into itinerant
electrons and vice versa,   different from a non-converting
mixture of mobile charged bosons and fermions \cite{ale3,and}.
 More recently this boson-fermion model has
been adopted for a description of  superfluidity of atomic
fermions  scattered into
 bound (molecular) states \cite{chio}.

Most studies of BFM  below its transition into a low-temperature
condensed phase applied a
 mean-field approximation (MFA), replacing  zero-momentum boson operators by c-numbers
 and neglecting the boson self-energy  in the density sum rule \cite{rob2,lee,ran,kos,lar,dam,mic,chio}.
  When the bare boson energy is well above the chemical potential, the BCS ground state
  was
  found  with bosons being only virtually excited \cite{lee,ran}.
  MFA led to a conclusion
 that BFM exhibits features compatible with BCS characteristics
 \cite{kos}, and describes a crossover from the BCS-like to local pair  behaviour
 \cite{mic}.  The transition was found  more mean-field-like than
 the usual Bose condensation, i.e. characterized by a relatively
 small value of $Gi$ \cite{lar}.

 At the same time
 our study of BFM \cite{ale}  beyond MFA
revealed a crucial effect of the boson self-energy on the normal
state boson spectral function and the transition temperature
$T_{c}$. Ref.\cite{ale} proved that the Cooper pairing of fermions
via virtual bosonic states  is impossible in any-dimensional BFM.
It occurs only simultaneously with the Bose-Einstein condensation
of real bosons in 3D BFM \cite{ale2,ren}, and vanishes  in 2D BFM
due to the absence of the Bose-Einstein condensation in two
dimensions \cite{ale2}. The origin of this simultaneous
condensation lies in a  softening of the boson mode at $T=T_c$
caused by its hybridization with fermions. The energy of
zero-momentum bosons  is renormalized down to \emph{zero}  at
$T=T_c$, no matter how weak the boson-fermion coupling and how
large the bare boson energy are. Bosons look like overdamped
diffusive modes, rather than quasiparticles in the long-wave limit
\cite{ale,ale2,cris} contrary to the conclusion of Ref.\cite{ran0}
that there is 'the onset of coherent free-particle-like motion of
the bosons' in this limit.  One can expect that the boson
self-energy should qualitatively modify the whole
 condensed phase of 3D BFM below $T_c$.

In this Letter  a closed set of
  equations  for   fermion  and
boson Green's functions (GFs) is derived taking into account
self-energy effects in the   condensed state of 3D BFM.  There
exist a boson \emph{pair} condensate along with the fermion Cooper
pair and the single-particle boson  condensate in the model.
Remarkably, the Gor'kov expansion \cite{gor2} of GFs in the
strength of the order parameter yields a zero  linear term at
\emph{any} temperature below $T_c$,
 and
 $Gi=\infty$.  It shows that the transition is not a
mean-field second order transition, and there is no crossover from
the BCS-like to a local pair  behaviour at any values of the
parameters of BFM.

The  Hamiltonian of BFM in an external magnetic field ${\bf B=
\nabla \times A}$ is defined as
\begin{eqnarray}
H&=&\int d{\bf r}\sum_s \psi _{s}^{\dagger }({\bf r})\hat{h}({\bf
r})\psi _{s}({\bf r}) +g[\phi({\bf r})\psi _{\uparrow }^{\dagger
}({\bf r})\psi _{\downarrow }^{\dagger }({\bf
r})+H.c.]\nonumber\\
&+&E_0\phi^{\dagger}({\bf r})\phi({\bf r}),
\end{eqnarray}
where $\psi_{s}({\bf r})$ and $\phi({\bf r})$ are fermionic and
bosonic fields, $s=\uparrow, \downarrow$ is the spin, $E_0$ is the
bare energy of bosons with respect to their chemical potential
$2\mu$,  $\hat{h}({\bf r)=}-[\nabla +ie{\bf A(r)]}^{2}/(2m)-\mu$
is the fermion kinetic energy operator , and $g$ is the
hybridization interaction converting a boson into two fermions and
vice versa. Here and further I take $\hbar=c=k_B=1$, and the
volume of the system $V=1$.

The Matsubara field operators, $Q=\exp (H\tau )Q({\bf r)}\exp
(-H\tau ), \bar{Q}=\exp(H\tau) Q^{\dagger}({\bf r)}\exp (-H\tau )$
($Q\equiv\psi_s, \phi$) evolve with the imaginary time
$-1/T\leqslant \tau \leqslant 1/T$ as

\begin{eqnarray}
-\frac{\partial \psi _{\uparrow }({\bf r},\tau )}{\partial \tau } &=&\hat{h}(%
{\bf r)}\psi _{\uparrow }({\bf r},\tau )+g \phi({\bf r},
\tau)\bar{\psi} _{\downarrow }({\bf r,}\tau ), \\
\frac{\partial \bar{\psi} _{\downarrow }({\bf r,}\tau )}{\partial
\tau } &=&\hat{h}^{\ast }({\bf r)}\bar{\psi} _{\downarrow }({\bf
r,}\tau )-g \bar{\phi}({\bf r},\tau)\psi _{\uparrow }({\bf r,}\tau
), \\ -\frac{\partial \phi ({\bf r,}\tau )}{\partial \tau} &=& E_0
\phi({\bf r,}\tau )+g\psi _{\downarrow }({\bf r,}\tau )\psi
_{\uparrow }({\bf r,}\tau ).
\end{eqnarray}
The  theory of the condensed state can be formulated with the
 normal and anomalous fermion GFs \cite{gor2}, $ {\cal
G}({\bf r,r}^{\prime },\tau)=- \langle T_{\tau
}\psi _{s}({\bf r},\tau )\bar{\psi} _{s}({\bf r}^{\prime }{\bf ,}%
0)\rangle$, ${\cal F}^{+}({\bf r,r}^{\prime },\tau)= \langle
T_{\tau }\bar{\psi} _{\downarrow }({\bf r,}\tau) \bar{\psi}
_{\uparrow }({\bf r}^{\prime
},0)\rangle$, respectively, where the operation $%
T_{\tau }$ performs the time ordering. Fermionic and bosonic
fields condense simalteneously \cite{ale}. Following Bogoliubov
\cite{bog} the bosonic  condensate is described by separating a
large matrix element $\phi_{0} ({\bf r})$ in $\phi({\bf r},\tau) $
as a number, while the remaining part $\tilde{\phi}({\bf r},\tau)$
describes a  supracondensate field, $ \phi ({\bf r},\tau)=\phi
_{0}({\bf r})+\tilde{\phi}({\bf r},\tau)$. Then using Eq.(4) one
obtains
\begin{equation}
g\phi_0({\bf r})= \Delta({\bf r})\equiv-{g^2\over{E_0}}{\cal
F}({\bf r,r},0+),
\end{equation}
where ${\cal F}({\bf r,r}^{\prime },\tau)= \langle T_{\tau }\psi
_{\downarrow }({\bf r,}\tau) \psi _{\uparrow } ({\bf r}^{\prime
},0)\rangle$. The equations for  GFs are obtained by using Eqs
(2,3,4) and the  diagrammatic technique \cite{abr} in the
framework of the non-crossing approximation \cite{ref}, as shown
in Fig.1 and Fig.2.

An important novel feature of BFM is a pairing of supracondensate
bosons, generated by their hybridization with the fermionic
condensate, as follows from the last diagram in Fig.2. Hence, one
has to introduce an $anomalous$ supracondensate boson GF, ${\cal
B}^{+}({\bf r,r}^{\prime },\tau)=\langle T_{\tau
}\bar{\tilde{\phi}}({\bf r,}\tau) \bar{\tilde {\phi}}({\bf
r}^{\prime },0)\rangle$ along with a normal boson GF, ${\cal
D}({\bf r,r}^{\prime },\tau)=- \langle T_{\tau
}\tilde{\phi} ({\bf r},\tau )\bar{\tilde{\phi}}({\bf r}^{\prime }{\bf ,}%
0)\rangle$.
\begin{figure}
\begin{center}

\includegraphics[angle=-90,width=0.57\textwidth]{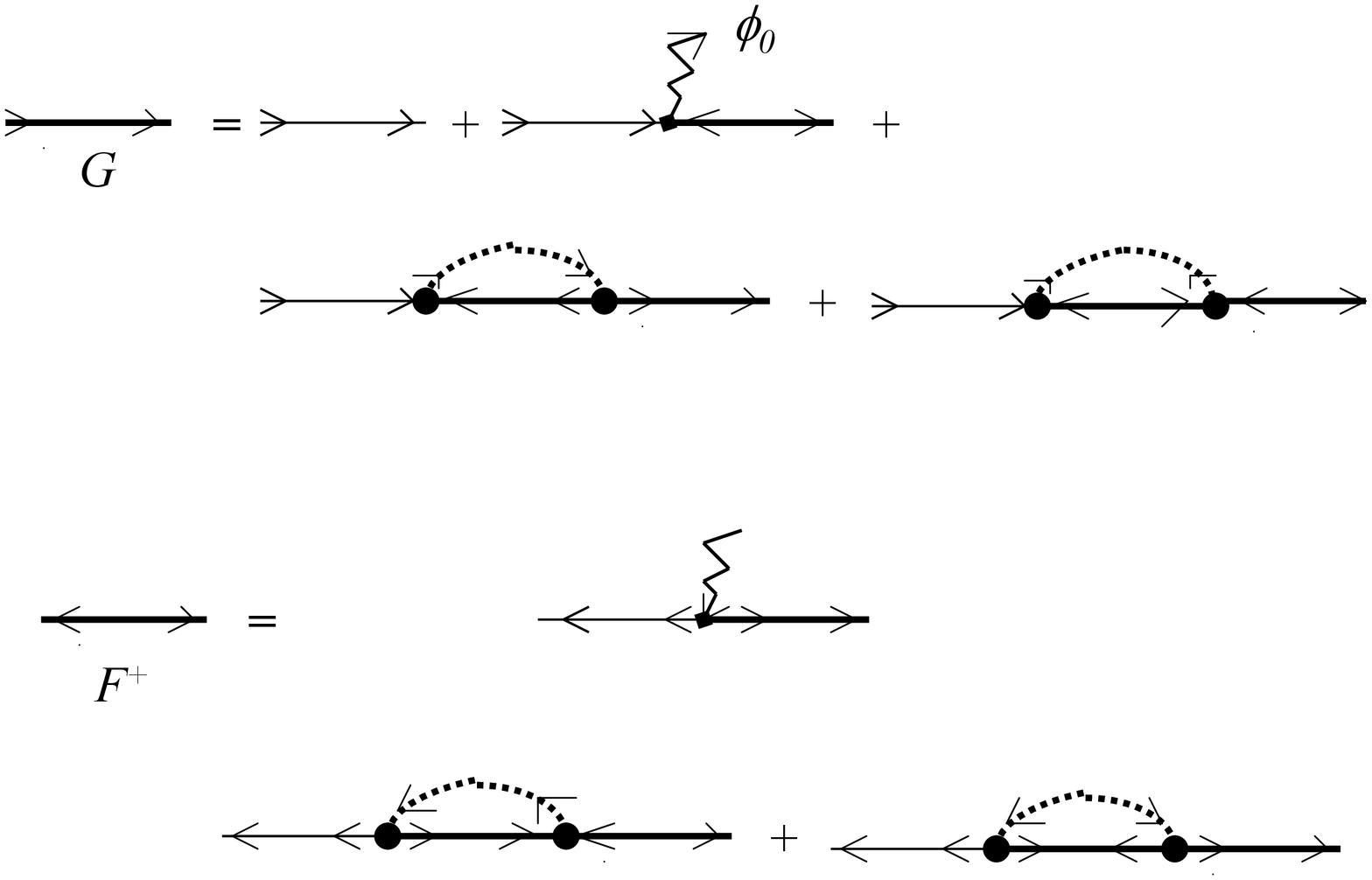}
\caption{Diagrams for the normal and anomalous fermion GFs.
Vertexes (dots) correspond to the hybridization interaction $g$,
and zig-zag arrows represent the single-particle Bose condensate
$\phi_0$.}
\end{center}
\end{figure}
The diagrams, Fig.1 and Fig.2, are transformed into analytical
equations for  time Fourier-components of the fermion GFs with the
Matsubara frequencies $\omega=\pi T (2n+1)$ ($n=0, \pm 1,\pm
2,...$) as
\begin{eqnarray}
&&[i\omega -\hat{h}({\bf r)}]{\cal G}_{\omega}({\bf r,r}^{\prime
})
 = \delta ({\bf r-r}^{\prime })-\Delta ({\bf r)}{\cal
F}_{\omega}^{+}({\bf r,r}^{\prime } ) \cr
 &-& g^2 T
\sum_{\omega^{\prime}} \int d{\bf x}{\cal
G}_{-\omega^{\prime}}({\bf x,r}){\cal
D}_{\omega-\omega^{\prime}}({\bf r,x}){\cal G}_{\omega}({\bf
x,r}^{\prime})
 \cr &-& g^2T
 \sum_{\omega^{\prime}} \int d{\bf
x}{\cal F}^{+}_{\omega'}({\bf r,x}){\cal
B}_{\omega+\omega^{\prime}} ({\bf r,x}){\cal F}^{+}_{\omega}({\bf
x,r^{\prime}}),
\end{eqnarray}

\begin{eqnarray}
&&[-i\omega -\hat{h}^{\ast}({\bf r)}]{\cal F}^{+}_{\omega}({\bf
r,r}^{\prime }) = \Delta^{\ast} ({\bf r)}{\cal G}_{\omega}({\bf
r,r}^{\prime } ) \cr
 &-& g^2 T
\sum_{\omega^{\prime}} \int d{\bf x}{\cal
G}_{\omega^{\prime}}({\bf r,x}){\cal
D}_{\omega^{\prime}-\omega}({\bf x,r}){\cal F}^{+}_{\omega}({\bf
x,r}^{\prime}) \cr &+& g^2T
 \sum_{\omega^{\prime}} \int d{\bf
x}{\cal F}_{-\omega'}({\bf r,x}){\cal
B}^{+}_{-\omega-\omega^{\prime}} ({\bf r,x}){\cal G}_{\omega}({\bf
x,r^{\prime}})
\end{eqnarray}
 and,
\begin{eqnarray}
&&(i\Omega -E_0) {\cal D}_{\Omega}({\bf r,r}^{\prime }) = \delta
({\bf r-r}^{\prime })\cr &-& g^2 T \sum_{\omega^{\prime}} \int
d{\bf x}{\cal G}_{\omega^{\prime}}({\bf r,x}){\cal
G}_{\Omega-\omega^{\prime}}({\bf r,x}){\cal D}_{\Omega}({\bf
x,r}^{\prime}) \cr &-& g^2T
 \sum_{\omega^{\prime}} \int d{\bf
x}{\cal F}_{\omega'}({\bf r,x}){\cal F}_{\Omega-\omega^{\prime}}
({\bf r,x}){\cal B}^{+}_{\Omega}({\bf x,r}^{\prime}),
\end{eqnarray}
\begin{eqnarray}
&&(-i\Omega -E_0){\cal B}^{+}_{\Omega}({\bf r,r}^{\prime }) = \cr
&&  g^2T
 \sum_{\omega^{\prime}} \int d{\bf
x}{\cal F}_{-\omega'}^{+}({\bf r,x}){\cal
F}_{-\Omega+\omega^{\prime}}^{+}({\bf r,x}){\cal D}_{\Omega}({\bf
x,r^{\prime}}) \cr &-& g^2 T \sum_{\omega^{\prime}} \int d{\bf
x}{\cal G}_{-\omega^{\prime}}({\bf x,r}){\cal
G}_{\omega^{\prime}-\Omega}({\bf x,r}){\cal B}^{+}_{\Omega}({\bf
x,r}^{\prime}).
\end{eqnarray}
for the boson GFs  with $\Omega=2\pi T n$, and ${\cal B}({\bf
r,r}^{\prime },\tau)=\langle T_{\tau }\tilde{\phi}({\bf r,}\tau)
\tilde {\phi}({\bf r}^{\prime },0)\rangle$

\begin{figure}
\begin{center}

\includegraphics[angle=-90,width=0.57\textwidth]{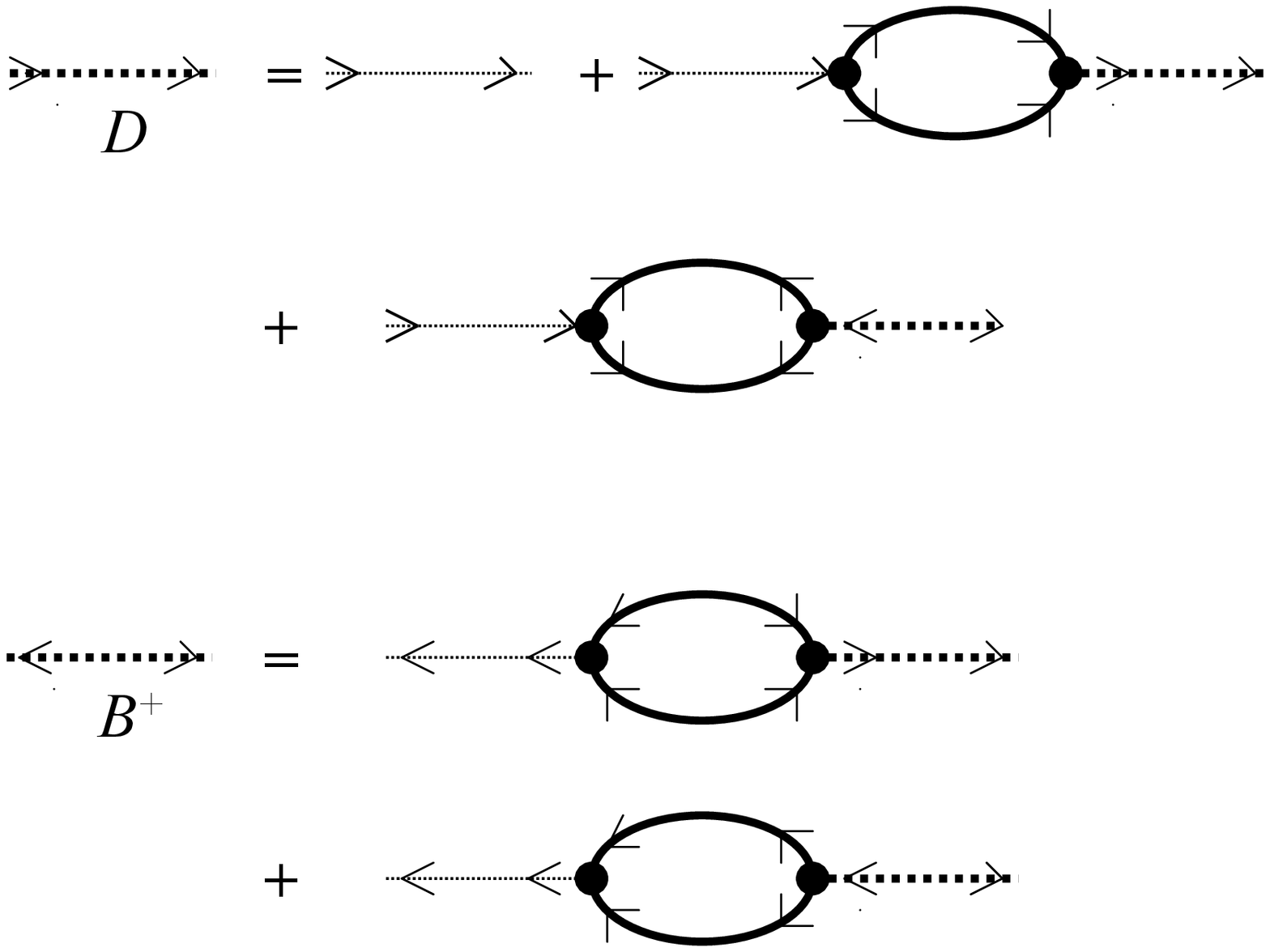}
 \caption{Diagrams for the supracondensate boson
 GFs. The Cooper-pairing of fermions  leads to the Cooper-pair-like boson condensate,
 described by the boson anomalous GF, ${\cal
B}^{+}$.}
\end{center}
\end{figure}

These equations can be formally solved in the homogeneous case
without the external field, ${\bf A}=0$. Transforming into the
momentum space yields GFs' time-space Fourier components as
\begin{eqnarray}
{\cal G}({\bf k},\omega) &=-&{\frac{i\tilde{\omega}^{\ast} +\xi
_{\bf k}}{ |i\tilde{\omega} -\xi _{\bf
k}|^2+|\tilde{\Delta}({\bf k}, \omega)|^{2}}},\\
{\cal F}^{+}({\bf k}, \omega) &=&{\frac{\tilde{\Delta}^{\ast}({\bf
k}, \omega)}{ |i\tilde{\omega} -\xi _{\bf
k}|^2+|\tilde{\Delta}({\bf k}, \omega)|^{2}}} ,
\end{eqnarray}
and
\begin{eqnarray}
{\cal D}({\bf q},\omega) &=&-{\frac{i\tilde{\Omega}^{\ast} +E_0}
{|i\tilde{\Omega} -E_0|^2+|\Gamma({\bf q}, \Omega)|^{2}}},
 \\
{\cal B}^{+}({\bf q}, \omega) &=&{\frac{\Gamma^{\ast}({\bf q},
\Omega) }{ |i\tilde{\Omega} -E_0|^2+|\Gamma({\bf q},
\Omega)|^{2}}},
\end{eqnarray}
where $\tilde{\omega}\equiv\omega+i\Sigma_f({\bf k}, \omega)$, $
\tilde{\Omega}\equiv\Omega+i\Sigma_b({\bf q}, \Omega)$.  The
fermionic order parameter, renormalised with respect to the
mean-field $\Delta$ due to the formation of the boson-pair
condensate, is given by
\begin{equation}
\tilde{\Delta}({\bf k}, \omega)=\Delta + g^2T
 \sum_{\omega^{\prime}}\int {d {\bf q}\over{2\pi^3}} {\cal F}^{+}({\bf k}-{\bf q},\omega'
){\cal B}({\bf q},\omega+\omega^{\prime}),
\end{equation}
and the boson-pair order parameter, generated by the hybridization
with the fermion Cooper pairs, is
\begin{equation}
 \Gamma({\bf
q}, \Omega)= g^2T \sum_{\omega^{\prime}}\int {d {\bf
k}\over{2\pi^3}} {\cal F}({\bf k},\omega'){\cal F}({\bf q}-{\bf
k},\Omega-\omega^{\prime}).
\end{equation}
  Hence,
there are three coupled condensates in the model described by the
 off-diagonal fields $g\phi_0$, $\tilde{\Delta}$, and
$\Gamma$, rather than two, as in MFA. At low temperatures all of
them have about the same magnitude, as the fermion and boson
self-energies,
\begin{equation}
\Sigma_f({\bf k}, \omega)=-g^2T
 \sum_{\omega^{\prime}}\int {d {\bf q}\over{2\pi^3}} {\cal G}({\bf q}-{\bf k},-\omega'){\cal
D}({\bf q},\omega-\omega^{\prime}),
\end{equation}
\begin{equation}
\Sigma_b({\bf q}, \Omega)=-g^2T
 \sum_{\omega^{\prime}}\int {d {\bf q}\over{2\pi^3}} {\cal G}({\bf k},\omega'){\cal
G}({\bf q}-{\bf k},\Omega-\omega^{\prime}),
\end{equation}
respectively.

 On the other hand, when the temperature is   close to $T_c$ (i.e.  $T_c-T \ll
 T_c$),
   the boson pair condensate  is weak compared with the single-particle boson and the Cooper pair condensates.
  Since
$\Gamma$, Eq.(15) is of the second order in $\Delta$, $\Gamma
\propto \Delta^{2}$, the anomalous boson GF can be neglected in
this temperature range, where $\Delta$ is small. The fermion
self-energy, Eq.(16) is a regular function of $\omega$ and ${\bf
k}$, so that it can be absorbed in the renormalized fermion band
dispersion. Then the fermion normal and anomalous GFs, Eqs.(10,11)
look like the familiar GFs  of the BCS theory, and one can apply
the Gor'kov expansion \cite{gor2}  in powers of $\Delta ({\bf r)}$
to describe the condensed phase of BFM in the magnetic field near
the transition.  Using Eq.(7) and Eq.(5) one obtains
 to the terms linear in $\Delta $
\begin{equation}
\Delta^{\ast} ({\bf r)} ={g^2\over{E_0}}T\sum_{\omega _{n}}\int d{\bf x}{\cal G}%
_{-\omega _{n}}^{(n)}({\bf x,r})\Delta ^{\ast }({\bf x)}{\cal
G}_{\omega _{n}}^{(n)}({\bf x,r}).
\end{equation}
The spatial variations of the vector potential  are small near the
transition.  If ${\bf A}({\bf r})$ varies slowly, the normal state
GF, ${\cal G}_{\omega }^{(n)}({\bf r,r}^{\prime })$ differs from
the zero-field normal state GF, ${\cal G}_{\omega
}^{(0)}(\bf{r-r^{\prime}} )$ only by a phase \cite{gor2} ${\cal
G}_{\omega }^{(n)}({\bf r,r}^{\prime })=\exp [-ie{\bf A(r)}\cdot
(\bf{r-r}^{\prime })]{\cal G}_{\omega }^{(0)}(\bf{r-r}^{\prime }
)$. Expanding all quantities  near the point ${\bf x=r}$ in
Eq.(18) up to the second order in ${\bf x-r} $ inclusive, one
obtains the linearised  equation for the fermionic order parameter
as
\begin{equation}
 \gamma [\nabla -2ie{\bf
A(r)]}^{2}\Delta({\bf r})=\alpha \Delta({\bf r}),
\end{equation}
where
\begin{equation}
\alpha = 1+{\Sigma_b(0,0)\over{E_0}}\approx 1-
{g^2N(0)\over{E_0}}\ln {\mu\over{T}},
\end{equation}
and $\gamma \approx 7\zeta (3)v_{F}^{2}g^2N(0)/(48\pi^2
T^{2}E_0)$.
 Here $v_F$ is the Fermi velocity, and $N(0)$ is the (renormalized)
 fermion density of states at the Fermi level.

In the framework of MFA one takes the bare boson energy   in
Eq.(20) as a temperature independent parameter, $E_0=g^2N(0)\ln
(\mu/T_c)$ \cite{lar}, or determines it from the conservation of
the total number of particles ( the density sum-rule) neglecting
the boson self-energy $\Sigma_b({\bf q}, \Omega)$
\cite{ran,dam,mic,chio}). Then Eq.(19) looks like the conventional
mean-field Ginzburg-Landau (GL) equation with a negative $\alpha
\propto T-T_c$, a relatively small fluctuation region $Gi$
\cite{lar}, and a finite $H_{c2}(T)$  \cite{dam}. As a result one
concludes that the phase transition is almost the conventional
BCS-like transition, at least at $E_0\gg T_c$ \cite{lee,lar}, and
BFM, Eq.(1) describes the crossover from the BCS-like to local
pair behaviour by tuning the parameters \cite{mic}. This
conclusion is incorrect. The main problem with MFA stems from the
density sum-rule, which determines the chemical potential of the
system and consequently the bare boson energy $E_{0}(T)$ as a
function of temperature,
\begin{equation}
-T\sum_{\Omega_n}e^{i\Omega _{n}\tau }\int {d{\bf q}\over{2\pi^3}}D({\bf q}%
,\Omega _{n})=n_{b}-\int d{\bf r} |\phi_0({\bf r})|^2.
\end{equation}
Here $\tau =+0$, $n_{b}=n-n_f$ is the number of bosons, $n $ is
the total number of particles, and $n_f$ is the number of
fermions. The  term of the sum in Eq.(21) with $\Omega_n=0$ is
given by the integral
\begin{equation}
T\int {d{\bf q}\over{2\pi^3}}{1\over{E_0+\Sigma_b({\bf q},0)}},
\end{equation}
where $\Sigma _{b}({\bf q},0)=\Sigma _{b}(0,0)+q^{2}/2M^{\ast }$
  for  a small ${\bf q}$  is calculated using Eq.(17) with the
  normal state fermion GF \cite{ale} (here $M^{\ast}$ is a constant). The integral converges, if and
only if $E_0\geqslant -\Sigma_b(0,0)$. This exact result means
that $\alpha \geqslant 0$ at any temperature.  In fact, this
coefficient is strictly zero in the Bose-condensed state, because
$\mu_b=-[E_0+\Sigma_b(0,0)]$ corresponds to the boson chemical
potential relative to the lower edge of the boson energy spectrum.
More generally, $\mu_b=0$ corresponds to the appearance of the
Goldstone-Bogoliubov mode due to a broken symmetry  below $T_c$.
As a result, the GL coefficient $\alpha (T)$, is \emph{zero}  at
\emph{any} temperature below $T_c$, $\alpha (T)\equiv 0$, and not
only at $T_c$ in the exact theory of BFM. On the other hand, MFA
 violates the density
sum-rule, predicting the wrong negative $\alpha(T)$ below $T_c$.

 There are  a few important physical consequences. Since $\alpha(T)=0$,
 the Levanyuk-Ginzburg parameter
 \cite{lev} is infinite. It means that the  phase
 transition is never a BCS-like second-order phase transition
 even at large $E_0$ and small $g$. In fact, the
 transition  is driven by the Bose-Einstein condensation of \emph{
 real} bosons with ${\bf q}=0$, which occur  due to the complete
 softening of their spectrum at  $T_c$ in 3D BFM.
 Remarkably, the conventional upper critical field, determined as the field, where a non-trivial
 solution of the \emph{linearised} Gor'kov equation (19)  occurs, is
 zero in BFM, $H_{c2}(T)=0$. It is not  a finite $H_{c2}(T)$
 found in Ref. \cite{dam} using MFA. In the homogeneous case $\Delta(T)$ should be determined from Eq.(21)
 rather than from the BCS-like equation (5), which
is actually an identity \cite{ale}, since $E_0=-\Sigma_b(0,0)$
below $T_c$. To get an insight
 into the magnetic properties of the condensed phase
one has to solve Eqs.(6-9, 14-17) and Eq.(21) keeping the
non-linear terms. Even
 at  temperatures well below $T_c$ the condensed state is fundamentally
 different from the MFA ground state, because of the pairing of
 bosons. The latter is similar to the Cooper-like pairing of
 supracondensate $^{4}He$ atoms \cite{pas}, proposed  as an explanation of the small density of the single-particle
 Bose condensate in superfluid Helium-4.   The pair-boson condensate
 should significantly
modify the thermodynamic
  properties of the condensed BFM
  compared with the MFA predictions.
   The common wisdom
 that at weak coupling  the boson-fermion model is adequately described by the BCS
 theory, is therefore negated by our theory.

I highly appreciate enlightening   discussions with A.F. Andreev,
L.P. Gor'kov, V.V. Kabanov, A.I. Larkin, and  A.P. Levanyuk, and
support by the Leverhulme Trust (UK) via Grant F/00261/H.


\begin{thebibliography}{90}
\bibitem{sim}
E. Simanek, Solid State Commun. {\bf 32}, 731 (1979).
\bibitem{tin}
C.S. Ting, D.N. Talwar and K.L. Ngai, Phys. Rev. Lett. {\bf 45},
1213 (1980).
\bibitem {ion}
 S.P. Ionov, Izv. AN SSSR Fiz {\bf
49}, 310 (1985).
\bibitem{rob} J. Ranninger and S. Robaszkiewicz, Physica B
(Amsterdam) {\bf 53}, 468 (1985). It was shown  that this
interpretation of the polaron-bipolaron crossover stemmed from a
misunderstanding of the polaron dynamics in the  Holstein model
 (Y.A. Firsov, V.V. Kabanov, E.K. Kudinov,  and A.S. Alexandrov,
  Phys. Rev. B {\bf 59}, 12132 (1999); A.S. Alexandrov, ibid {\bf 61}, 12315 (2000)).
\bibitem{rob2}
S. Robaszkiewicz, R. Micnas, and J. Ranninger, Phys. Rev. B {\bf
36}, 180 (1987).
\bibitem {eli}
G.M. Eliashberg, Pis'ma Zh. Eksp. Teor. Fiz. (Prilozh.) {\bf 46},
94 (1987) (JETP Lett. Suppl. {\bf 46}, S81  (1987)).
\bibitem{lee} R.
Friedberg and T.D. Lee, Phys. Rev. B{\bf 40}, 6745 (1989); R.
Friedberg, T.D. Lee, and H.C. Ren, Phys.Rev. B{\bf 42}, 4122
(1990).
\bibitem{ran}
R. Micnas, J. Ranninger, and S. Robaszkiewicz, Rev. Mod. Phys.
{\bf 62}, 113 (1990); J. Ranninger and J.M. Robin, Physica C{\bf
235}, 279 (1995).
\bibitem{ran0} J. Ranninger, J.M. Robin, and M.
Eschrig, Phys. Rev. Lett. {\bf 74}, 4027 (1995).
\bibitem{kos} T. Kostyrko and J.
Ranninger, Phys. Rev. B {\bf 54}, 13105 (1996).

\bibitem{lar} V.B. Geshkenbein, L.B. Ioffe, and A.I Larkin,
Phys. Rev.B {\bf 55}, 3173 (1997).
\bibitem{ale}  A. S. Alexandrov, J. Phys.: Condens. Matter {\bf 8}, 6923
(1996).
\bibitem{ale2}
 A.S. Alexandrov, Physica C{\bf 274}, 237 (1997); ibid {\bf %
316}, 239 (1999).
\bibitem{ren}  R. Friedberg, H. C. Ren and O. Tchernyshyov, J. Phys.:
Condens. Matter {\bf 10}, 3089 (1998); A. S. Alexandrov, ibid,
3093.
\bibitem{cris} I. Grosu, C. Blaga, and M. Crisan, J.
Supercond. {\bf 13}, 459 (2000).
\bibitem{gor} L. P. Gor'kov, J.
Supercond. {\bf 13}, 765 (2000)

\bibitem{dam}  T. Doma\'{n}ski, Phys. Rev. B{\bf 66},
134512 (2002);  T. Doma\'{n}ski, M. M. Ma\'{s}ka and M.
Mierzejewski, Phys. Rev. B 67, 134507 (2003).
\bibitem{mic} R. Micnas, S.
Robaszkiewicz, and A. Bussmann-Holder, in \emph{Highlights in
condensed matter physics} (eds. A. Avella et al., AIP Conference
Proceedings, Melville, New York ) {\bf 695}, 230 (2003) and
references therein.

\bibitem{ale3} A.S. Aleksandrov
(Alexandrov) and A.B. Khmelinin, Fiz. Tverd. Tela (Leningrad) {\bf
28}, 3403 (1986) (Sov. Phys. Solid State {\bf 28}, 1915 (1986)).
\bibitem{and} A.F. Andreev, Pis'ma Zh. Eksp. Teor. Fiz. {\bf 79}, 100 (2004).
\bibitem{chio}  M.L. Chiofalo, S.J.J.M.F. Kokkelmans, J.N.
Milstein, and M.J. Holland, Phys. Rev. Lett. {\bf 88}, 090402
(2002).
\bibitem{gor2} L.P. Gor'kov, Zh. Eksp. Teor.
Fiz. {\bf 34}, 735 (1958) (Sov. Phys.-JETP {\bf 7}, 505 (1958));
ibid {\bf 36}, 1918 (1959) (Sov. Phys.-JETP {\bf 9}, 1364 (1958))

\bibitem{bog}N.N. Bogoliubov, Izv. Acad. Sci. (USSR) {\bf 11}, 77
(1947).
\bibitem{abr} A.A. Abrikosov, L.P. Gor'kov, and I.E.
Dzyaloshiskii, \emph{Methods of  Quantum Field Theory in
Statistical Physics} (Prentice-Hall, Englewood Cliffs, N.J.,
(1963)).
\bibitem{ref} The crossing diagrams are small  as $g/\mu \ll 1$
\cite{ale}. Taking them  as well as  hard-core and other dynamic
interactions into account change none of our conclusions.

\bibitem{lev}  L.D. Landau and E.M. Lifshitz, \emph{Statistical
Physics, 3rd Edition,  Part 1} (eds. E.M. Lifshitz and L.P.
Pitaevskii, Pergamon Press, Oxford- New York (1980)), page 476.
\bibitem{pas} E.A.
Pashitskii, S.V. Mashkevich, and S.I. Vilchynskyy, Phis. Rev.
Lett. {\bf 89}, 075301 (2002).



\end{thebibliography}
 \end{document}